\documentclass[twocolumn,showpacs,showkeys,preprintnumbers,amsmath,amssymb]{revtex4}

\usepackage{graphicx}
\usepackage{dcolumn}
\usepackage{bm}

\begin{document}

\title{Magnetic ground state of coupled edge-sharing CuO$_2$ spin-chains}

\author{U.~Schwingenschl\"ogl}
\author{C.~Schuster}
\affiliation{Institut f\"ur Physik, Universit\"at Augsburg, 86135 Augsburg, Germany}

\date{\today}

\begin{abstract}
By means of density functional theory, we investigate the magnetic
ground state of edge-sharing CuO$_2$ spin-chains, as found in the
(La,Ca,Sr)$_{14}$Cu$_{24}$O$_{41}$ system, for instance. Our data
rely on spin-polarized electronic structure calculations including
onsite interaction (LDA+U)
and an effective model for the interchain coupling. Strong doping
dependence of the magnetic order is characteristic for edge-sharing
CuO$_2$ spin-chains. We determine the ground state magnetic structure
as function of the spin-chain filling and quantify the competing
exchange interactions.
\end{abstract}

\pacs{71.10.Pm, 71.20.-b, 75.25.+z, 75.30.Et}

\keywords{density functional theory, electronic structure, magnetic chain}

\maketitle

Doping of magnetic systems nowadays attracts great interest due to a large number
of exotic effects. 3D antiferromagnetic order, for example, is easily
destroyed by either quantum or thermodynamical fluctuations, and by
any kind of impurity. The study of doped 2D antiferromagnets \cite{Vojta}
has been triggered by the phase diagrams of high-temperature superconductors.
Upon doping, they show a transition from an antiferromagnetic to a
superconducting state. On the other hand, 1D copper oxide compounds
are an active field of research, too. From the theoretical point
of view, the magnetic properties of undoped two-leg Cu$_2$O$_3$
ladders are explained in terms of a Heisenberg model with cyclic
ring exchange \cite{nunner02}. Doping effects are investigated in Refs.\
\cite{jeckelmann98,nishimoto02}. Antiferromagnetic CuO chains with
180$^\circ$ Cu-O-Cu bond angles, as found in Sr$_2$CuO$_3$
\cite{Motoyama}, for example, can be understood by conformal field
theory \cite{Affleck}. When a few Cu ions are replaced by non-magnetic
ions, like Zn, the system is well described by the Heisenberg model
with removed spins \cite{Eggert}. Edge-sharing CuO$_2$ chains with
90$^\circ$ Cu-O-Cu bond angles, on the contrary, are frustrated due
to competition between ferromagnetic and antiferromagnetic exchange
\cite{Mizuno_98,Tornow,drechsler2007}. In this context, both charge
and spin order of highly doped systems are not understood by far.

The (Sr,Ca,La)$_{14}$Cu$_{24}$O$_{41}$ system of isostructural
materials is prototypical for doping dependent charge and magnetic
ordering in a partially filled spin-chain, i.e.\ a chain with less
than one spin per site. The compounds have been subject of intensive
research in recent years, mainly due to a rich phase diagram and a
close relation to cuprate superconductors. The incommensurate
crystal structures consist of planes of quasi 1D CuO$_2$ chains
running along the crystallographic $c$-axis, stacked alternately with
planes of two-leg Cu$_2$O$_3$ ladders. The lattice constants of these
subsystems satisfy $10c_{\rm chain}\approx7c_{\rm ladder}$
\cite{ukei94,structure}. Materials with different Sr/Ca
composition are isostructural with only minor modifications of the bond
lengths and angles \cite{Ohta,Matsuda97,Isobe}. A relative shift
between adjacent chains parallel to the chain axis distinguishes
Sr and Ca rich compounds. In Ca$_{13.6}$Sr$_{0.4}$Cu$_{24}$O$_{41}$
the shift amounts to half the intrachain Cu-Cu distance, while it is
only 30\% of this distance for Sr$_{14}$Cu$_{24}$O$_{41}$. For each
Sr/Ca composition the Cu valence is 2.25, where Cu$^{2+}$ ions ($S=1/2$)
accumulate on the ladders. In contrast, most of the 6 Cu$^{3+}$
ions per unit cell, which form $S=0$ Zhang-Rice singlets (so-called
holes), are located on the chains. Optical conductivity and x-ray
absorption data suggest that substitution of Ca for Sr induces
a transfer of holes back into the ladders \cite{Osafune97,Nuecker00}. 
Consequently, an average Cu valence of 2.50 is assumed for the
chain sites in Ca rich samples, giving rise to a half filled chain
with one spin per two sites. On La substitution the intrinsic doping
decreases gradually, reaching the undoped state with a nominal Cu valence of
2.00 for La$_6$Ca$_8$Cu$_{24}$O$_{41}$.

By chain geometry, competing ferromagnetic nearest-neighbour and
antiferromagnetic next-nearest neighbour coupling is expected for
edge-sharing CuO$_2$ spin-chains. The magnetic phase diagram
comprises at least 3 phases upon doping, i.e.\ variation of the
number of holes in the spin-chain. In the case of Sr$_{14}$Cu$_{24}$O$_{41}$
the CuO$_2$ chains are non-magnetic with a spin gap of about 130\,K,
which can be explained in a dimer picture \cite{Matsuda96}. Using a
cluster calculation, Gelle\'e and Lepetit \cite{Gellee} have shown
that the ground state is very complex, with trimer contributions.
A dimer ground state is found by exact diagonalization of a Heisenberg
model with nearest and next-nearest neighbour interaction \cite{klingeler06},
in accordance with band structure data \cite{srcuo2}. On the contrary,
antiferromagnetic intrachain ordering has been reported for Ca rich
samples \cite{Nagata,Isobe}. Despite a huge number of investigations,
most details of the antiferromagnetic order are still a matter of
controversy. Even worse, in Sr$_{13}$LaCu$_{24}$O$_{41}$ spin dimers
evolve instead of the antiferromagnetism expected for half filled chains
\cite{Gerber2005}. Reflecting Cu-O-Cu bond angles of $\approx90^\circ$,
ferromagnetism finally is realized in slightly doped spin-chains of La
rich systems \cite{Carter,Matsuda96b}, which are captured by a 2D
classical Heisenberg model with defects \cite{Selke}.

Spin-chain systems are strongly affected by both subtleties
of the crystal structure and the electron-electron interaction.
In the case of half filled spin-chains \cite{csus07}
the ground state magnetic structure has been established in
\cite{magstr}. However, the dependence of the exchange coupling on
the filling $n$ (where n = number of spins / number of sites)
is an open question, since the coupling over a hole competes with
the coupling over an interstitial spin. In order to obtain a comprehensive
picture of doped spin-chains, we thus clarify the details of the
transition from antiferromagnetic to ferromagnetic coupling on
reducing the hole count. To be more specific, we study the onset
of ferromagnetism at $n=2/3$ and the full spin chain at $n=1$
for typical Cu-Cu bond lengths of 2.75\,\AA\ (nearest neighbour) and
5.8\,\AA\ (next-nearest neighbour).

The following results are based on density functional theory
and the generalized gradient approximation. We apply the WIEN2k
program package, a state-of-the-art full-potential code
with a mixed lapw and apw+lo basis \cite{wien2k}. For all the spin configurations
under investigation, the charge density is represented by at least 70,000 plane waves and
the Brillouin zone integration uses a {\bf k}-mesh comprising at least 18
{\bf k}-points in the irreducible wedge. While Cu $3p$ and O $2s$
orbitals are treated as semi-core states, the valence states
contain Cu $3d$, $4s$, $4p$ and O $2p$, $3s$ orbitals. To achieve a
mixed valence in a unit cell with both Cu$^{3+}$ and Cu$^{2+}$
ions, we start the calculation from the following setup: One electron is added to
each oxygen, thus preventing charge transfer off the copper, and
restricted to a Watson sphere of radius 1.3\,\AA. Moreover, a $3d^{10}$
configuration is used for the Cu$^{3+}$ ions and a $3d^{10}4s^1$
configuration for the Cu$^{2+}$ ions.

\begin{table}[t]
\begin{tabular}{l|c|c|c|c|}
{\bf staggered}&\;\;fm\;\;&\;afm\;&4k${\rm_F}$/afm&cluster\\\hline
${\rm U}=0$\,Ryd:\;\;${\rm E-E_0}$ (mRyd)&45&---&15&0.3\\\hline
${\rm U}=0.3$\,Ryd:\;\;${\rm E-E_0}$ (mRyd)&19&21&27&---
\end{tabular}\\[0.5cm]
\begin{tabular}{l|c|c|c|c|}
{\bf symmetric}&\;\;fm\;\;&\;afm\;&4k${\rm_F}$/afm&cluster\\\hline
${\rm U}=0$\,Ryd:\;\;${\rm E-E_0}$ (mRyd)&$-6$&---&64&1\\\hline
${\rm U}=0.3$\,Ryd:\;\;${\rm E-E_0}$ (mRyd)&18&27&66&---
\end{tabular}
\vspace*{0.2cm}
\caption{\rm Total energy gain (per Cu atom) with respect to a
non-magnetic calculation, for both staggered and symmetrically
aligned spin-chains. The fm, afm, 4k${\rm_F}$/afm, and cluster
spin patterns are compared to each other \cite{magstr}. The filling
of the spin-chain is $n=1/2$.}
\label{tab1}
\end{table}

\begin{figure}[b]
\includegraphics[width=0.35\textwidth,clip]{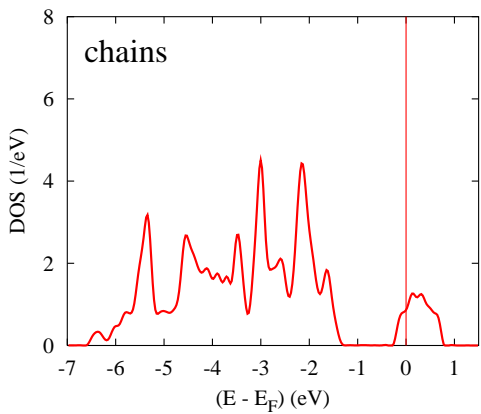}
\caption{Density of states (DOS) for the chain subsystem of Sr$_{14}$Cu$_{24}$O$_{41}$.
\cite{Schuster_epjb}}
\label{figdos}
\end{figure}

From a structural point of view, the spin-chain materials
(La,Ca,Sr)$_{14}$Cu$_{24}$O$_{41}$ consist of 3 subsystems: The
CuO$_2$ spin-chains, Cu$_2$O$_3$ ladders, and interstitial electron
donor ions. As interaction between the subsystems is very small, an
effective model of coupled spin-chains, capturing structural
prerequisites of the exchange interaction, may be used to investigate
the compounds \cite{srcuo2}. The density of states (DOS) of the CuO$_2$
chains shows a structure of width $\approx5.2$\,eV at binding energies
larger than 2\,eV and a second structure of width 1\,eV right at the
Fermi level. As expected for transition metal oxides \cite{prb07,ve0405,adp04},
the states are due to bonding between Cu $3d_{xz}$ and O $2p_x/p_z$
orbitals. Characteristic Cu $3d_{xz}$ bands at the Fermi energy
dominate the electronic and magnetic properties \cite{Schuster_epjb}.
A tight-binding fit reveals strong next-nearest neighbour hopping,
which is rather independent of the interchain coupling. Nearest
neighbour hopping, in contrast, can be suppressed by relative shifts
between the CuO$_2$ chains. In that case the transverse hopping, a
measure of the 2D coupling, increases \cite{srcuo2}.

\begin{figure}[t]
\includegraphics[width=0.35\textwidth,clip]{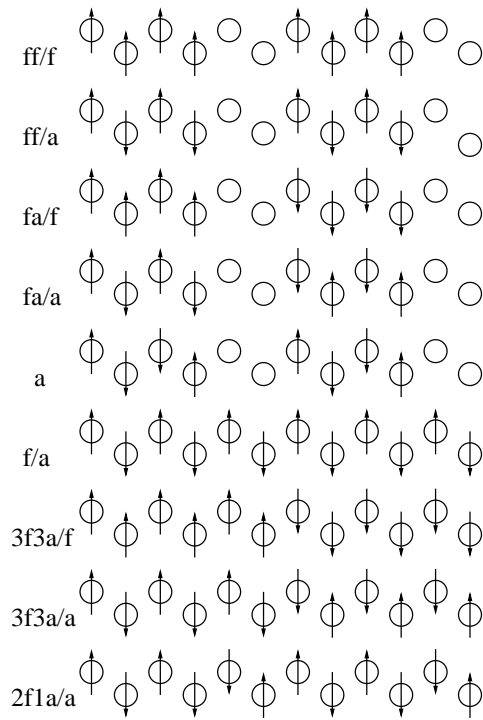}
\caption{Nomenclature of spin patterns on coupled chains.}
\label{fign23}
\end{figure}

\begin{table*}[t]
\begin{tabular}{l|c|c|c|c|c|}
&ff/f&ff/a$\to$f/a&fa/f$\to$3f3a/f&fa/a$\to$3f3a/a&a\\\hline
${\rm U}=0$\,Ryd:\;\;${\rm E-E_0}$ (mRyd)&5&12&3&5&9\\[0.5cm]
&ff/f&f/a&3f3a/f&3f3a/a&a$\to$2f1a/a\\\hline
${\rm U}=0.3$\,Ryd:\;\;${\rm E-E_0}$ (mRyd)&27&32&30&30&28
\end{tabular}
\vspace*{0.2cm}
\caption{\rm Total energy gain (per Cu atom) with respect to a
non-magnetic calculation for the spin-chain filling factor $n=2/3$.
Several spin patterns are compared to each other, see fig.\ \ref{fign23}
for the nomenclature.}
\label{tab2}
\end{table*}
\begin{table}[t]
\begin{tabular}{l|c|c|c|}
&f/f&f/a&a\\\hline
${\rm U}=0$\,Ryd:\;\;${\rm E-E_0}$ (mRyd)&\;$-16$\;&\;45\;&\;105\;\\\hline
${\rm U}=0.3$\,Ryd:\;\;${\rm E-E_0}$ (mRyd)&30&28&30
\end{tabular}
\vspace*{0.5cm}
\caption{\rm Total energy gain (per Cu atom) with respect to a
non-magnetic calculation for a full spin-chain ($n=1$). The f/f, f/a,
and a spin patterns are compared to each other.}
\label{tab3}
\end{table}

For $n=1/2$, the local density approximation (LDA), not
accounting for the onsite interaction, yields a ferromagnetic ground
state for staggered spin-chains \cite{magstr}. In contrast,
antiferromagnetic spin order and 4k${\rm_F}$-periodic charge order,
complemented by antiferromagnetic interchain coupling, are observed
for symmetric spin-chains, compare the data in Table \ref{tab1}.
In this case, the ferromagnetic solution is even unstable. Including
local electron-electron interaction at the correlated Cu $3d$ orbitals,
an LDA+U treatment reveals an antiferromagnetic ground state in both
arrangements, see Table \ref{tab1}. A relative shift between
the CuO$_2$ chains promotes antiferromagnetism more effectively than
a local interaction. In total, next-nearest neighbour coupling
over a hole site controls the magnetism in a half filled
spin-chain. The combination of electron-electron repulsion and
structural frustration therefore gives rise to a 4k$_F$ spin pattern.
Due to a weak interchain coupling of about 1\,mRyd, the magnetic
next-nearest neighbour interaction energy is $54\pm2$\,mRyd
per Cu atom within the chain.

Turning to a lower hole doping, we start with the filling $n=2/3$ and
afterwards consider the full spin-chain, given by $n=1$. For $n=2/3$, we
have 4 Cu$^{2+}$ ions per 6 lattice sites on the coupled chains.
The supercell entering the band structure calculation therefore contains
12 inequivalent Cu sites. We set up the start configuration as follows:
Within a chain, a ferro- or antiferromagnetic two spin cluster is
coupled to the next such cluster over a hole site. Experimental data
rather point at antiferromagnetic than ferromagnetic coupling between
adjacent chains \cite{Matsuda96b}. However, as the details of the
exchange interaction are not known, we study both possibilities for
comparison. This leads to altogether five configurations, which we
systematically call ff/f, fa/f, ff/a, fa/a and a. For clarity, these
configurations are depicted in fig.\ \ref{fign23}. In each case, we
calculate the total energy and compare it to the findings of a
spin-degenerate calculation, yielding the energy gain due to the
exchange interaction.

The fully ferro- (ff/f) and antiferromagnetic (a) cluster spin configurations
in the following serve as a reference frame for the discussion.
Without local electron-electron interaction, only these two patterns
are found to be stable. The magnetic moments are $\approx0.35\mu_{\rm B}$
on the Cu$^{2+}$-sites, where moments connected to two holes are
slightly larger than those connected to a single hole. Ferromagnetic
clusters with antiferromagnetic intrachain coupling (fa/f and fa/a)
transform into modified clusters without intermediate holes, therefore
into a more homogeneous charge distribution. According to the initial
interaction, we obtain a 3f3a/f or 3f3a/a spin pattern, as depicted in
fig.\ \ref{fign23}. Ferromagnetic intrachain coupling in combination
with antiferromagnetic interchain coupling (ff/a) results in a homogeneous
spin-chain (fa). For the spin patterns under consideration, table \ref{tab2}
compares the total energy gain due to the magnetic exchange interaction.

When we account for a local electron-electron interaction
${\rm U}=0.3$\,Ryd by means of the LDA+U method, the spin patterns
ff/f, 3f3a/f, 3f3a/a and f/a remain stable, where the magnetic
moments increase slightly to about 0.5\,$\mu_B$. Only the fully
antiferromagnetic cluster configuration develops a more homogeneous
charge distribution given by the spin pattern 2f1a/a, as shown in
fig.\ \ref{fign23}. We obtain the largest total energy gain (with
respect to the non-magnetic state) for the f/a spin pattern, amounting
to 32\,mRyd per Cu site. The magnetic ground state of edge-sharing
CuO$_2$ chains with filling $n=2/3$ therefore is given by fully polarized
ferromagnetic spin-chains and antiferromagnetic interchain coupling.
Remarkably, the electronic correlations do not alter the magnetic ground
state, see the ${\rm U}=0$\,Ryd data in table \ref{tab2}. As compared
to the coupling in a half filled spin-chain, ferromagnetic nearest
neighbour interaction here becomes dominant as the electron-electron
repulsion supports a homogeneous charge distribution. Whereas we find
the onset of ferromagnetism already for $n=2/3$, experiments on
La-rich systems suggest a nominal filling of about $n=0.7$. We ascribe
this discrepancy to insufficient knowledge about the effective chain
hole count in the real materials. By the data given in table
\ref{tab2}, we conclude that the energy gain due to the interchain
magnetic coupling is at most 1\,mRyd per lattice site. In addition,
comparison of the f/a to the 2f1a/a data allows us to estimate an
energy difference of 1.3\,mRyd between the ferro- and antiferromagnetic
coupling of neighbouring spins. The combination of the ff/f data with
the half filled spin chain results in an intrachain nearest neighbour
exchange interaction energy of $27\pm2$\,mRyd per Cu atom. Therefore,
the next-nearest neighbour coupling is found to be almost exactly twice
as strong as the nearest neighbour coupling.

For the full spin-chain ($n=1$), we address ferro- and antiferromagnetically
coupled CuO$_2$ chains with a ferromagnetic intrachain coupling
(f/f and f/a), and a frustrated antiferromagnet (a). For all three
configurations the band structure calculation converges with magnetic
moments of about $0.4\,\mu_{\rm B}$. The antiferromagnet has the lowest energy, both
for ${\rm U=0}$\,Ryd and a moderate local electron-electron interaction
(of ${\rm U}=0.3$\,Ryd), see table \ref{tab3}. In the latter case, the magnetic
moments increase to 0.6 $\mu_{\rm B}$ for the antiferromagnet.
In contrast to the doped compounds, a local interaction now counteracts
the spin ordering. By an energy difference of less than $0.2$\,mRyd per
lattice site, ferro- and antiferromagnetic interaction between nearest
neighbour spins result in almost degenerate electronic states.
On the contrary, the energy gain due to the interchain coupling is
larger than determined for the spin-chain fillings $n=1/2$ and $n=2/3$.

In conclusion, we have investigated the magnetic coupling in
edge-sharing CuO$_2$ chains with $\approx90^\circ$ Cu-O-Cu bond angles,
at doping levels of $n=1/2$, $n=2/3$, and $n=1$. By means of LDA+U
electronic structure calculations, we have studied the total energy
for various spin patterns. Our data shows that accounting for both
the chain geometry and the electron-electron interaction \cite{schnack05}
is essential for obtaining a comprehensive picture of the magnetism,
though electronic correlation does not alter the magnetic ground state.
For symmetric CuO$_2$ chains and filling $n=1/2$ the latter is
antiferromagnetic with a 4k${\rm_F}$ charge order along the chain axis.
Adjacent chains are coupled antiferromagnetically, too. This
result particularly applies to the coupled CuO$_2$ chains of Ca-rich
spin-chain materials (Sr,Ca)$_{14}$Cu$_{24}$O$_{41}$. La-rich systems
with spin-chain filling $n=2/3$ develope a ferromagnetic intrachain
coupling since Coulomb repulsion supports a homogeneous distribution
of the charge. The interchain coupling remains antiferromagnetic. When
the filling is complete ($n=1$), finally a frustrated antiferromagnet
is found, which fact reflects strong electronic correlations. Under the
assumption that modelling in terms of a Hubbard model is appropriate,
antiferromagnetism supports intrachain hopping, which is excluded
for ferromagnetic spin-chains. By Nagaoka's theorem \cite{nagaoka66}, a
single hole is expected to turn the system into a ferromagnetic state,
in agreement with results for La$_5$Ca$_9$Cu$_{24}$O$_{41}$.
However, the experimental observation of antiferromagnetic spin-chains
in undoped La$_6$Ca$_8$Cu$_{24}$O$_{41}$ is missing so far.

\begin{acknowledgments}
We gratefully acknowledge various fruitful discussions with V.\ Eyert
and T.\ Kopp, as well as financial support by the Deutsche
Forschungsgemeinschaft (SFB 484).
\end{acknowledgments}

\end{document}